\documentclass[11pt]{article}
\usepackage{graphicx}
\usepackage{cite}

\usepackage{amsthm}

\begin{document}

%
%
\title{\Large\bf
On the process dependent nuclear $k_\perp$ broadening effect}

\author{  Andreas Sch\"{a}fer and Jian Zhou
 \\[0.3cm]
{\normalsize\it Institut f\"{u}r Theoretische Physik,Universit\"{a}t
 Regensburg, Regensburg, Germany}
}

\maketitle

\begin{abstract}
We study the process dependent nuclear $k_\perp$ broadening effect by employing the transverse momentum
dependent(TMD) factorization approach in combination with the Mclerran-Venugopalan(MV) model.
More specifically, we investigate how the parton transverse momentum distributions are affected by the process
dependent gauge links in cold nuclear matter.
In particular, our analysis also applies to the polarized cases including
the nuclear quark Boer-Mulders function and the linearly polarized gluon distribution.
Our main focus is on the nuclear TMDs at intermediate or large x.
\end{abstract}

%
%
\section{Introduction}
\noindent
The detailed understanding of the properties of hot and cold nuclear matter is one of the topical problems of QCD,
in particular in connection with high energy heavy-ion experiments at RHIC, LHC and the future EIC.
Initial/final state multiple parton re-scattering in a large nucleus
plays an important role in revealing the properties of cold nuclear matter,
as it leads to  various physical effects, such as transverse momentum broadening of the propagating parton,
 parton energy loss due to induced gluon bremsstrahlung, and  nuclear dependence of  azimuthal asymmetries.

Much efforts have been devoted to the study of transverse momentum broadening in eA and pA collisions.
A number of different approaches developed to describe this phenomenon
 were formulated within  different theoretical frameworks, such as
the eikonal approximation~\cite{Bodwin:1981fv},
 twist-4 collinear factorization~\cite{Luo:1992fz} and the resummation of
higher twist contributions~\cite{Fries:2002mu,Majumder:2007hx},
 dipole approach~\cite{Dolejsi:1993iw,Johnson:2000dm},
 the BDMPS formalism~\cite{Baier:1996sk,Wu:2011kc,Liou:2013qya},
  diagrammatic Glauber multiple scattering~\cite{Gyulassy:2002yv},
 color glass condensate(CGC) effective theory~\cite{Dumitru:2001jn,Kharzeev:2003wz},
 transverse momentum dependent factorization(TMD)~\cite{Liang:2008vz},
 and soft collinear effective theory(SCET)~\cite{Idilbi:2008vm,D'Eramo:2010ak,Ovanesyan:2011xy,Benzke:2012sz}.
The energy dependence of nuclear $k_\perp$ broadening has also been investigated in Refs.~\cite{CasalderreySolana:2007sw,Mueller:2012bn}.

Within the TMD factorization approach~\cite{Collins:1981uw,Ji:2004wu}, we identified
 the gauge link appearing in the matrix element definition for nuclear TMDs as the main source of leading nuclear effects~\cite{Liang:2008vz}.
The formalism we developed in~\cite{Liang:2008vz} was used to study $k_\perp$ broadening as well as
the nuclear dependence of azimuthal asymmetries~\cite{Gao:2010mj,Song:2010pf,Gao:2011mf}
in semi-inclusive DIS(SIDIS) off a large nucleus.
In SIDIS  nuclear TMDs contain  a future pointing gauge link describing  the final state interactions,
 while a past pointing gauge link shows up in the nuclear TMDs associated with  the Drell-Yan process in pA collisions
 due to   initial state interactions. The contributions to $k_\perp$ broadening
from the future pointing and the past pointing gauge links are identical as this is a T-even observable.
In the processes involving more complicated color flow,  parton transverse momentum distributions can be
affected by  both initial and final state interactions, and thus could significantly
differ from these in SIDIS and DY processes. The initial/final state interactions leading to $k_\perp$ broadening
in eA and pA collisions can be encoded in the various process dependent gauge links~\cite{Bomhof:2004aw}.
The purpose of this paper is to investigate the process dependent
nuclear TMDs at intermediate or large $x$ following our general method described in~\cite{Liang:2008vz}.
As a byproduct, one can readily deduce  $k_\perp$ broadening simply by computing the $k_\perp^2$ moment of nuclear TMDs.

As a matter of fact, the process dependent nuclear TMDs at small $x$ have been studied
in both the unpolarized~\cite{Xiao:2010sp,Dominguez:2010xd,Dominguez:2011wm} and polarized cases~\cite{Metz:2011wb,Akcakaya:2012si}.
In general, small $x$ nuclear TMDs associated with different hard scattering  processes
recover the same well known perturbative tail $1/k_\perp^2$ in the dilute medium limit,
 while they could differ significantly in the dense medium case where initial/final multiple re-scattering plays a
 more important role for parton transverse momentum spectra.
One remarkable example is the difference between two widely used small x gluon TMDs in saturation physics:
 the Weizs\"{a}cker-Williams(WW) gluon distributions and the dipole gluon distribution.
As pointed out in~\cite{Dominguez:2010xd,Dominguez:2011wm},
 the WW gluon distribution contains a future pointing or past pointing gauge link in
the adjoint representation while the dipole distribution contains a closed loop gauge link.
Both gluon distributions can be probed in  different hard scattering processes~\cite{Dominguez:2010xd,Dominguez:2011wm} .

In the present paper, we are aiming to extend the analyses~\cite{Xiao:2010sp,Dominguez:2010xd,Dominguez:2011wm,Metz:2011wb,Akcakaya:2012si}
 to the intermediate or large $x$ region.
At small $x$,  TMDs are perturbatively calculable due to the presence of a semi-hard scale (the so called
saturation scale), generated dynamically in high energy scattering. In contrast, nuclear TMDs at intermediate
or large x can not be computed perturbatively. However, for the same reason, one can calculate contributions from
initial/final state interactions encoded in process dependent gauge links in the MV model. By doing so, we are able to
express nuclear TMDs as  the convolution of the corresponding nucleon
ones and process dependent small $x$ gluon distributions.
$k_\perp$ broadening is obtained as byproduct from the relation between the nuclear TMD and the nucleon TMD in a specific
hard scattering process.

At this point, we would like to briefly comment on the factorization properties of the relevant processes.
Factorization in terms of TMDs containing process dependent gauge links is often referred to as generalized
transverse momentum dependent(GTMD) factorization~\cite{Bomhof:2004aw}.
In the framework of GTMD factorization, the modified gauge links
are obtained by resumming longitudinally polarized gluons into
parton correlation functions on each nucleon side separately.
However, recent work has shown that it is impossible to do so for di-jet production in pp collisions
because the initial/final state interaction will not allow a separation of
gauge links into the matrix elements of the various TMDs associated with each incoming proton.
This has been explicitly illustrated by a concrete counter-example in Ref.~\cite{Rogers:2010dm}.
In pA collisions, if one only takes into account the interaction between the active partons and
the background gluon field inside a large nucleus while
neglecting the longitudinal gluons attached to the proton side, the type of graph
(for example Fig.11 in~\cite{Rogers:2010dm})
which can produce a violation of generalized TMD-factorization disappears.
After neglecting the extra gluon attachment on the proton side,  multiple gluon re-scattering between
the hard part and the nucleus can be resummed to all orders in the form
of a process dependent gauge link. As a result, the predictive power of the theory is partly restored in pA collisions.

We also noticed that the process dependent $k_\perp$ broadening effect has been studied within the
twist-4 collinear factorization approach~\cite{Kang:2008us,Kang:2011bp,Xing:2012ii}.
The fact that the twist-4 collinear approach can be applied in the intermediate
and large x region allows us to directly compare our formalism with the high-twist approach.
It is shown that two approaches yield identical physical results for $k_\perp$ broadening in different processes
provided that the  saturation scale and the twist-four quark gluon correlation functions
are parameterized in a similar manner.

The paper is organized as follows: in Sec. 2, we review our general method developed in~\cite{Liang:2008vz}
and apply a modified version to compute the nuclear enhancement of the transverse momentum imbalance for quark pair production in eA collisions,
 and Drell-Yan di-lepton production in pA collisions. In addition, we establish relations between
 nuclear quark Boer-Mulders distributions and nucleon ones in SIDIS and Drell-Yan using the same approach.
 In Sec. 3, we study  nuclear $k_\perp$ broadening for photon-jet production and quark pair production
 in pA collisions, respectively.  In Sec. 4, we compare our results
 with those obtained in the collinear twist-4 approach and discuss the phenomenology implications.
 We conclude the paper and summarize our work in Sec. 5.

\section{Nuclear  TMDs in SIDIS and Drell-Yan }
\noindent
In our original work~\cite{Liang:2008vz},
multiple gluon correlations from the gauge link appearing in the definition of  nuclear quark TMDs are
 reduced to  products of nucleon small x gluon
distribution in the so-called maximal two-gluon correlation approximation. From such expression,
we obtained  nuclear TMD as a convolution of a Gaussian distribution and a nucleon quark TMD.
The width of the Gaussian is given by the gluon distribution density in the nuclear medium.
However, the evaluation of Wilson lines in the MV model~\cite{McLerran:1993ni} has already reached a rather sophisticated level.
In the present work, we will, therefore, compute the contribution from gauge links using the MV model instead of the maximal
two-gluon correlation approximation.
Both the unpolarized nuclear TMDs and  the polarized TMDs (quark or gluon Boer-Mulders distributions)
can be treated in the same framework.
The resulting expression of the gauge link contribution in  SIDIS
computed in the MV model no longer has a Gaussian form
 once the finite nuclear matter size effect are taken into account.
 We start our derivation with the nuclear TMDs containing a simple future pointing or past pointing gauge link.

\subsection{ Semi-inclusive DIS scattering off a large nucleus}
In this sub-section, we investigate how the out-going quark transverse momentum spectrum is affected by
 final state interactions encoded in the future pointing gauge link in the SIDIS process.
The explicit relations between the nucleon quark TMDs and the corresponding nuclear ones are established.

Our starting point is the operator definition of quark TMDs in an unpolarized nucleon,
\begin{eqnarray}
{\cal M}_N(x,\vec k_\perp) & = & \int \frac{d r^- d^2 r_{\perp}}{(2\pi)^3 }
\, e^{ixP^+r^- - i\vec{k}_{\perp} \cdot \vec{r}_\perp} \langle
N | \bar \psi (y^-,y_\perp) \,  U^{[+]} \psi(r^-+ y^-, r_{\perp}+y_\perp )   |N \rangle
\nonumber\\ & = & \frac{1}{2} f_{1,DIS}(x,k_\perp)p \!\!\!/ +\frac{1}{2k_\perp}h^\perp_{1,DIS}(x,k_\perp) \sigma^{\mu\nu}k_\mu p_\nu
\,,
\end{eqnarray}
where  $k_\perp$ is defined as $k_\perp\equiv |\vec{k}_{\perp}|$, and $p^\mu$ is commonly defined light cone vector.
The average over coordinate $y$ is implied.
 $|N \rangle$ represents the nucleon state.
 $f_{1,DIS}(x,k_\perp)$ is the normal unpolarized quark distribution function containing
 a future pointing gauge link which arises from the final state interaction in the semi-inclusive DIS process.
The second parton distribution  $h^\perp_{1,DIS}(x,k_\perp)$ is commonly  referred to as quark Boer-Mulders function~\cite{Boer:1997nt}.
Note that our convention  for the quark Boer-Mulders function differs from the literatures~\cite{Zhou:2008fb,Zhou:2009jm,Zhou:2009rp}
 by a factor $k_\perp/M_N$, where $M_N$ is  target mass.
Roughly speaking, the quark Boer-Mulders function describes the strength of the correlation between the quark transverse polarization and its transverse momentum.
Color gauge invariance is ensured by two (future-pointing) gauge links in the fundamental representation,
\begin{equation}
U^{[+]} = \mathcal{P} \, e^{-ig \int_{y^-}^{\infty} d \zeta^- A^+(\zeta^-, \, y_{\perp})} \,
\mathcal{P} \, e^{-ig \int_\infty^{r^-+y^-}
d\zeta^-  A^+(\zeta^- , \, r_\perp+y_\perp )} \, .
\end{equation}
We choose to work in the covariant gauge in which $A^+$  is the dominant component.
The transverse pieces of the gauge link are suppressed and will be neglected in the derivation presented below~\cite{Belitsky:2002sm}.

Generally speaking, the longitudinal polarized gluons building up the gauge link may carry arbitrary collinear nucleon momentum fractions , since the
gluon pole is not pinched.
However, in this paper, we only focus on the contribution from small x gluons, which can be computed perturbatively due to the presence of a semi-hard scale
 $Q_s$. One might expect that the contribution to $k_\perp$ broadening
from gluons with large (or intermediate) longitudinal momentum fraction is suppressed
as compared to that from  small x gluons because of the high
gluon number density at small x and the larger transverse momentum carried by these gluons.

Following a standard procedure (see, e.g., Ref.\cite{Iancu:2002xk} for an overview),
one first solves the classical Yang-Mills equation and obtains,
\begin{eqnarray}
A^+_a(y^-,y_\perp)=-\frac{1}{\nabla_\perp^2} \rho_a(y^-,y_\perp) \ ,
\end{eqnarray}
where $a$ is color index.
We proceed by inserting this solution into the gauge link and averaging over the color sources $\rho(x^-,x_\perp)$
 with the Gaussian distribution $W [\rho ]$~\cite{McLerran:1993ni}.
\begin{equation}
W_A[\rho]= \exp \Big\{-\frac{1}{2} \int d^3 y
\frac{ \rho_a(y^-,y_\perp) \rho_a(y^-,y_\perp)}{\lambda_A(y^-)} \Big\} \ ,
\end{equation}
where $\lambda_A(y^-)$ is the density of the color charges at a given $y^-$.
With this ansatz for the distribution of color sources, the most elementary correlator is given by,
\begin{eqnarray}
 \langle A^+_a(y^-,y_\perp) A^+_b(r^-,r_\perp) \rangle
&=&\delta_{ab} \delta(y^--r^-) \Gamma_A(y_\perp-r_\perp) \lambda_A(y^-) \
\nonumber \\
\Gamma_A(k_\perp)&\equiv&\frac{1}{k_\perp^4} \ .
\end{eqnarray}
By repeatedly using this elementary correlator, one evaluates a pair of Wilson lines stretching from $ y^-+R^-$ to infinity
 with $R^-$ being the nucleon radius.
\begin{eqnarray}
&&
\left  \langle \left [ \mathcal{P} \, e^{-ig \int_{y^-+R^-}^{\infty} d \zeta^- A^+(\zeta^-, \, y_{\perp})} \,
\mathcal{P} \, e^{-ig \int_\infty^{y^-+R^-}
d\zeta^-  A^+(r_\perp+y_\perp, \, \zeta^-)} \right ]_{ab} \right \rangle
\nonumber \\ && =
\exp \bigg\{ - C_F \Theta(r_\perp^2)
\int_{R^-+y^-}^{\infty} d\zeta^- \, \lambda_A(\zeta^-) \bigg\} \delta_{ab}  \, .
\end{eqnarray}
where $\Theta(r_\perp^2) \equiv g^2 [ \Gamma_A(0_\perp) -\Gamma_A(r_\perp)] \simeq g^2 \frac{r_\perp^2}{16\pi} {\rm ln} \frac{1}{r_\perp^2\Lambda_{QCD}^2}$.
Note that these two Wilson lines are connected in  color space at infinity. The resulting expression is a
production of the unitary color matrix  and an exponential. This procedure is illustrated in Fig.1.
\begin{figure}[t]
\begin{center}
\includegraphics[width=16cm]{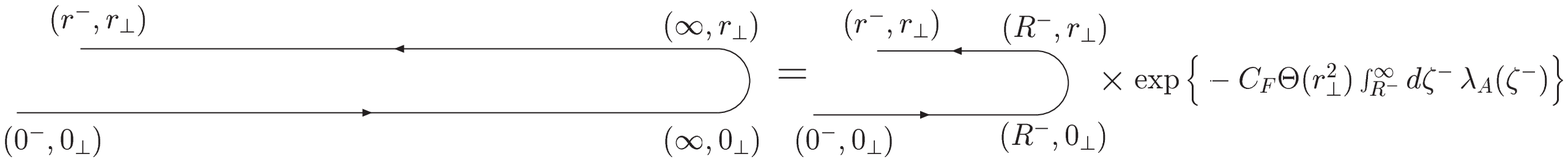}
\end{center}
\vskip -0.4cm \caption{The ordinary future pointing gauge link is reduced to a short one
by evaluating part of the gauge link stretching from $R^-$ to $\infty$ in the MV model, where $R^-$ is the radius of a nucleon.} \label{fig1}
\end{figure}

Inserting this expression into the  matrix element ${\cal M}_N $, one obtains,
\begin{eqnarray}
{\cal M}_N(x,\vec k_\perp)  &=&\int \frac{d r^- d^2 r_{\perp}}{(2\pi)^3 }
\, e^{ixP^+r^- - i\vec{k}_{\perp} \cdot \vec{r}_\perp}
 \langle N | \bar \psi (y^-,y_\perp) \, \bar U^{[+]} \, \psi(r^-+y^-, r_{\perp}+y_\perp )   |N \rangle
\nonumber \\ && \times
\exp \bigg\{ - C_F \Theta(r_\perp^2)
\int_{R^-+y^-}^{\infty} d\zeta^- \, \lambda_A(\zeta^-) \bigg\} \,.
\label{exp}
\end{eqnarray}
Here, the gauge link $\bar U^{[+]}$ is a short one and defined as,
\begin{eqnarray}
 \bar U^{[+]} = \mathcal{P} \, e^{-ig \int_{y^-}^{R^-+y^-} d \zeta^- A^+(\zeta^-, \, y_{\perp})} \,
 \mathcal{P} \, e^{-ig \int_{R^-+y^-}^{r^-+y^-} d \zeta^- A^+(\zeta^-, \, r_{\perp}+y_\perp)} \ .
\end{eqnarray}
which consists of two short Wilson lines connected in color space at point $R^-+y^-$.
Apparently, the density of the color sources outside a nucleon is
zero: $\int_{R^-+y^-}^{\infty} d\zeta^- \, \lambda_A(\zeta^-)=0$.
Such vanishing contribution from gauge link outside a nucleon has also been clearly seen
in the lattice calculation~\cite{Musch:2011er}.
As a result, the exponential factor in the Eq.\ref{exp} becomes unity. The
TMD  correlator is reduced to,
\begin{eqnarray}
{\cal M}_N(x,\vec k_\perp) &=&\int \frac{d r^- d^2 r_{\perp}}{(2\pi)^3 }
\, e^{ixP^+r^- - i\vec{k}_{\perp} \cdot \vec{r}_\perp}
 \langle N | \bar \psi (0^-,0\perp) \, \bar U^{[+]} \, \psi(r^-, r_{\perp} )   |N \rangle
 \nonumber\\ & = & \frac{1}{2} f_{1,DIS}(x,k_\perp)p \!\!\!/ +\frac{1}{2 k_\perp}h^\perp_{1,DIS}(x,k_\perp) \sigma^{\mu\nu}k_\mu p_\nu
\,,
\end{eqnarray}
where we have shifted the coordinate $y$ to zero using  translation invariance.

Note that the above derivation applies to both nuclear and nucleon targets and that all results have the
same form. The only difference is that for a large nucleus target, the struck nucleon is surrounded by cold nuclear matter,
such that the density of color sources outside the struck nucleon is no longer zero.
\begin{eqnarray}
{\cal M}_A(x,\vec k_\perp)  &=&\int \frac{d r^- d^2 r_{\perp}}{(2\pi)^3 }
\, e^{ixP^+r^- - i\vec{k}_{\perp} \cdot \vec{r}_\perp}
 \langle A | \bar \psi (y^-,y_\perp) \, \bar U^{[+]} \, \psi(r^-+y^-, r_{\perp}+y_\perp )   |A \rangle
\nonumber \\ && \times
\exp \bigg\{ - C_F \Theta(r_\perp^2)
\int_{R^-+y^-}^{\infty} d\zeta^- \, \lambda_A(\zeta^-) \bigg\} \,
 \nonumber\\ & = & \frac{1}{2} \textbf{f}_{1,DIS}(x,k_\perp)p \!\!\!/ +\frac{1}{2 k_\perp}\textbf{h}^\perp_{1,DIS}(x,k_\perp) \sigma^{\mu\nu}k_\mu p_\nu
\,,
\end{eqnarray}
where $\textbf{f}_{1,DIS}$ and $\textbf{h}_{1,DIS}^\perp$ denote the unpolarized quark  and quark Boer-Mulders TMD distributions inside a large nucleus respectively.
 To proceed further, we make two assumptions:

1) we neglect
the correlation between different nucleons and assume the large nucleus as a weakly bound,
\begin{eqnarray}
\langle A | \bar \psi (y^-,y_\perp) \, \bar U^{[+]} \, \psi(r^-+y^-, r_{\perp}+y_\perp )   |A \rangle
= \langle N | \bar \psi (0^-,0_\perp) \, \bar U^{[+]} \, \psi(r^-, r_{\perp} )   |N \rangle \int dy^- d^2y_\perp \rho_N^A(y) \ ,
\end{eqnarray}
where $|N>$ is understood as the nucleon state averaged over protons and neutrons inside a large nucleus,
 and $\rho_N^A(y)$ is the spatial nucleon density  normalized to the atomic number $\cal A$;

2) we further describe  the large nucleus as a homogenous system of nucleons and color sources,
\begin{eqnarray}
 \rho_N^A(y)= \rho_N^A(0)  \ \  , \ \lambda_A(\zeta^-)=\lambda_A(0^-)  \, .
\end{eqnarray}
Implementing these two approximations, one has,
\begin{eqnarray}
{\cal M}_A(x,\vec k_\perp)  &=& \int \frac{d r^- d^2 r_{\perp}}{(2\pi)^3 }
\, e^{ixP^+r^- - i\vec{k}_{\perp} \cdot \vec{r}_\perp}
 \langle N | \bar \psi (0^-,0_\perp) \, \bar U^{[+]} \, \psi(r^-, r_{\perp} )   |N \rangle
\nonumber \\ && \times  \int dy^- d^2y_\perp \,  \rho_N^A(y^-)
\exp \bigg\{ - C_F \Theta(r_\perp^2)
\int_{R^-+y^-}^{\infty} d\zeta^- \, \lambda_A(\zeta^-) \bigg\} \,
\nonumber \\ &\approx &
 {\cal A} \int \frac{d r^- d^2 r_{\perp}}{(2\pi)^3 }
\, e^{ixP^+r^- - i\vec{k}_{\perp} \cdot \vec{r}_\perp}
 \langle N | \bar \psi (0^-,0_\perp) \, \bar U^{[+]} \, \psi(r^-, r_{\perp} )   |N \rangle
 \frac{1-e^{-\frac{r_\perp^2 Q_{s,q}^2}{4}}}{r_\perp^2 Q_{s,q}^2/4}
 \label{nTMD}
\end{eqnarray}
In the second step of the above equation,  the approximation
 $\int_{R^-+y^-}^{\infty} d\zeta^- \, \lambda_A(\zeta^-)\approx \int_{y^-}^{\infty} d\zeta^- \, \lambda_A(\zeta^-)$
 valid for a large nucleus target has been used.
The quark saturation momentum $Q_{s,q}$ is given by
$Q_{s,q}^2=\alpha_sC_F {\rm ln}\frac{1}{r_\perp^2 \Lambda_{QCD}^2}\int_{-\infty}^\infty d\zeta^- \, \lambda_A(\zeta^-)$.

Eq.\ref{nTMD}  can be reexpressed in  momentum space as,
\begin{eqnarray}
{\cal M}_A(x,\vec k_\perp) = {\cal A } \int d^2 l_\perp  {\cal M}_N(x,\vec l_\perp) {\cal F}_{DIS}(|\vec k_\perp-\vec l_\perp|) \ .
\label{moment}
\end{eqnarray}
Here ${\cal F}_{DIS}(|\vec k_\perp-\vec l_\perp|)$ is given by,
\begin{eqnarray}
 {\cal F}_{DIS}(|\vec k_\perp-\vec l_\perp|)=\int \frac{ d^2 r_{\perp}}{(2\pi)^2 }\, e^{ - i(\vec k_{\perp}-\vec l_\perp) \cdot \vec r_\perp} \,
4 \frac{1-e^{-\frac{r_\perp^2 Q_{s,q}^2}{4}}}{r_\perp^2 Q_{s,q}^2} \ ,
\end{eqnarray}
and is normalized to 1:  $\int d^2l_\perp {\cal F}_{DIS}(l_\perp) =1$.
When $Q_{s,q}^2$ is  small,
the fact that $ {\cal F}_{DIS}(|\vec k_\perp-\vec l_\perp| )\simeq \delta^2(\vec k_\perp-\vec l_\perp) $
allows us to recover the usual nucleon TMDs.
Inserting the decomposition of the matrix correlator into Eq.\ref{moment},  it is easy to derive that,
\begin{eqnarray}
\textbf{f}_{1,DIS}(x,k_\perp)&=& {\cal A } \int d^2 l_\perp  \, f_{1,DIS}(x,l_\perp) {\cal F}_{DIS}(|\vec k_\perp-\vec l_\perp|)
\label{f1}
   \\
\textbf{h}_{1,DIS}^\perp(x,k_\perp)&=&{\cal A } \int d^2 l_\perp \, \left ( \hat k_\perp \cdot \hat l_\perp \right ) \,
h_{1,DIS}^\perp(x,l_\perp) {\cal F}_{DIS}(|\vec k_\perp-\vec l_\perp|) \ .
\label{h1perp}
\end{eqnarray}
The unit vectors $\hat k_\perp$ and $\hat l_\perp$ are defined as $\hat k_\perp\equiv \vec k_\perp/k_\perp$ and  $\hat l_\perp\equiv \vec l_\perp/l_\perp$, respectively.
 From these relations between  nucleon TMDs and nuclear TMDs, one  finds,
\begin{eqnarray}
\int d^2 k_\perp \, \textbf{f}_{1,DIS}(x,k_\perp)&=& {\cal A } \int d^2 l_\perp  \, f_{1,DIS}(x,l_\perp)={\cal A }f_{1}(x)
  \\
\int d^2 k_\perp \, k_\perp^2 \textbf{f}_{1,DIS}(x,k_\perp)&=& {\cal A } \int d^2 l_\perp  \,  l_\perp^2 f_{1,DIS}(x,l_\perp)
+\frac{1}{2} Q_{s,q}^2{\cal A } f_{1}(x)
  \\
\int d^2 k_\perp \,  k_\perp\, \textbf{h}_{1,DIS}^\perp(x,k_\perp)&=& {\cal A } \int d^2  l_\perp \, l_\perp \,
h_{1,DIS}^\perp(x,l_\perp)=-2\pi M_N {\cal A } T_F^{(\sigma)}(x,x)  \ .
\label{boer}
\end{eqnarray}
$f_{1}(x)$ is the normal integrated unpolarized quark distribution of a nucleon,
while $T_F^{(\sigma)}(x,x)$(convention used in Refs.~\cite{Zhou:2008fb,Zhou:2009jm}) is the
twist-3 quark gluon correlation function  inside an unpolarized nucleon.  In the last step of Eq.\ref{boer}, we used the well known relation between
the moment of the Boer-Mulders function and $T_F^{(\sigma)}(x,x)$~\cite{Boer:2003cm}. It turns out that this relation is not affected by the cold nuclear medium.
Moreover, for the approximations made in this paper, the unpolarized quarks inside a large nucleus are re-distributed in  transverse momentum space
while the total probability to find a quark carrying a certain longitudinal momentum fraction $x$ remains unchanged.
To be more specific, the quark transverse momentum distribution becomes broader and the transverse momentum broadening squared is $Q_{s,q}^2/2$.

In the semi-hard region where $k_\perp^2$ is of the order $Q_{s,q}^2$ , Eq.\ref{f1} and Eq.\ref{h1perp} can be simplified by dropping the terms suppressed by  powers
$<l_\perp^2>/Q_{s,q}^2$ where$<l_\perp^2>$ is the average squared parton intrinsic transverse momentum inside a nucleon,
\begin{eqnarray}
\textbf{f}_{1,DIS}(x,k_\perp)&\simeq& {\cal A } f_{1}(x) {\cal F}_{DIS}(k_\perp)
   \\
\textbf{h}_{1,DIS}^\perp(x,k_\perp)&\simeq&  {\cal A }2\pi M_N T_F^{(\sigma)}(x,x)
\frac{1}{2 } \frac{\partial {\cal F}_{DIS}(k_\perp)}{\partial k_\perp}  \ ,
\end{eqnarray}
which is the main result of this section. We believe that the main feature of multiple scattering in the cold nuclear matter
has been captured in the above equations though  various approximations were made in deriving them.

We conclude this subsection by emphasizing again the important point that the derivation presented above is only valid for nuclear TMDs at intermediate $x$ or large $x$.
In our calculation, it was critical to assume that the hard scattering takes place locally inside a nucleon, i.e., $r^-\leq R^-$.
This is in sharp contrast to the dipole model where the quark antiquark pair coherently interacts with the whole nucleus.

\subsection{The Drell-Yan process in pA collisions}
We now turn to  $k_\perp$ broadening  in Drell-Yan process.
In a widely used hybrid approach (for a review, see~\cite{JalilianMarian:2005jf}),
 one utilizes  ordinary integrated parton distributions for the dilute projectile proton, while
the  transverse momentum carried by a parton coming from a nucleus is left unintegrated.
We adopted the same strategy when dealing with pA collisions throughout this paper.
As a consequence, at lowest order,
the obtained  transverse momentum spectrum of the produced virtual photon is directly related to the $k_\perp^2$ moment of the
nuclear quark TMDs.

Quark TMDs appear in Drell-Yan differential cross sections contain a past-pointing gauge link,
\begin{equation}
U^{[-]} = \mathcal{P} \, e^{-ig \int_{y^-}^{-\infty} d \zeta^- A^+(\zeta^-, \, y_{\perp})} \,
\mathcal{P} \, e^{-ig \int_{-\infty}^{r^-+y^-}
d\zeta^-  A^+(\zeta^-, \, r_\perp+y_\perp )} \,.
\end{equation}
Following the procedure outlined in the previous sub-section, one obtains the same relations between nucleon quark TMDs and nuclear quark TMDs,
\begin{eqnarray}
\textbf{f}_{1,DY}(x,k_\perp)&=& {\cal A } \int d^2 l_\perp  \, f_{1,DY}(x,l_\perp) {\cal F}_{DY}(|\vec k_\perp-\vec l_\perp|)
   \\
\textbf{h}_{1,DY}^\perp(x,k_\perp)&=&{\cal A } \int d^2 l_\perp \, \left ( \hat k_\perp \cdot \hat l_\perp \right ) \,
h_{1,DY}^\perp(x,l_\perp) {\cal F}_{DY}(|\vec k_\perp-\vec l_\perp|) \ ,
\end{eqnarray}
with
\begin{eqnarray}
{\cal F}_{DY}(|\vec k_\perp-\vec l_\perp|) ={\cal F}_{DIS}(|\vec k_\perp-\vec l_\perp|).
\end{eqnarray}
Similarly, a shorter past-pointing gauge link emerges in the matrix element definition for $f_{1,DY}(x,l_\perp)$
and $h_{1,DY}^\perp(x,l_\perp)$,
\begin{eqnarray}
 \bar U^{[-]} = \mathcal{P} \, e^{-ig \int_{0}^{-R^-} d \zeta^- A^+(\zeta^-, \,  0_{\perp})} \,
 \mathcal{P} \, e^{-ig \int_{-R^-}^{r^-} d \zeta^- A^+(\zeta^-, \, r_{\perp})} \ .
\end{eqnarray}
In the above formula, the coordinate y has again been shifted to zero by translation invariance.
With this shorter gauge link, using time reversal and parity invariance, one may readily deduce,
\begin{eqnarray}
f_{1,DY}(x,l_\perp)=f_{1,DIS}(x,l_\perp)  \ \ \ \
 h_{1,DY}^\perp(x,l_\perp)= -h_{1,DIS}^\perp(x,l_\perp) \ ,
\end{eqnarray}
and therefore,
\begin{eqnarray}
\textbf{f}_{1,DY}(x,l_\perp)=\textbf{f}_{1,DIS}(x,l_\perp)  \ \ \ \
 \textbf{h}_{1,DY}^\perp(x,l_\perp)=- \textbf{h}_{1,DIS}^\perp(x,l_\perp) \ .
\end{eqnarray}
We notice that the unique universality property of the T-odd distribution $\textbf{h}_{1}^\perp$~\cite{Brodsky:2002cx,Collins:2002kn}
is preserved under our manipulation of gauge links. In the end, we would like to mention that the transverse momentum broadening
 for virtual photon produced  in pA collisions is also parameterized by $Q_{s,q}^2/2$.

\subsection{Heavy quark pair production in eA collisions}
We study the nuclear broadening of heavy quark-antiquark pair momentum imbalance in eA collisions.
Heavy quark production in eA collisions is initiated by the gluon channel,
\begin{eqnarray}
\gamma^*+g \rightarrow Q+\bar Q \, .
\end{eqnarray}
The transverse momentum imbalance of a quark-antiquark pair is defined as
\begin{eqnarray}
\vec k_\perp=\vec p_{1\perp}+\vec p_{2\perp}
\end{eqnarray}
where $\vec p_{1\perp}$ and $\vec p_{2\perp}$ are the transverse momenta of produced quark and antiquark, respectively.
In TMD factorization and at leading order,
$\vec k_\perp$ is identical to the transverse momentum carried by the incoming gluon simply because of  momentum conservation.
The matrix element definition for nuclear gluon TMDs  in SIDIS  is given in
~\cite{Mulders:2000sh,Anselmino:2005sh},
\begin{eqnarray} \label{e:ww}
{\cal M}_{A}^{ij}(x,\vec k_\perp) & = & \int \frac{d r^- d^2 r_{\perp}}{(2\pi)^3 P^+}
e^{ixP^+r^- - i\vec k_{\perp} \cdot \, \vec r_\perp}
\langle A | F^{+i}(y^-,y_\perp) \, \tilde U^{[+]}  \, L_{y} \,
F^{+j}( r^- + y^-, r_{\perp} + y_{\perp}) |A \rangle
\nonumber\\ & = &
\frac{\delta_{\perp}^{ij}}{2} \, x \textbf{G}_{DIS}(x, k_\perp) +
\bigg(\hat k_\perp^i \hat k_\perp^j - \frac{1}{2} \delta_{\perp}^{ij} \bigg)
x \textbf{h}^{\perp g}_{1,DIS}(x, k_\perp) \,,
\end{eqnarray}
where  $\tilde U^{[+]} $ is the future pointing gauge link in the adjoint representation.
$\textbf{G}_{DIS}$ and $\textbf{h}^{\perp g}_{1,DIS}$ stand for the unpolarized gluon TMD  and linearly polarized gluon TMD  respectively.
$\textbf{h}^{\perp g}_{1,DIS}$ is the only polarization dependent gluon TMD for an unpolarized nucleon/nucleus, and
therefore may be considered as the counterpart of the quark Boer-Mulders function. However,
in contrast to the later, $\textbf{h}^{\perp g}_{1,DIS}$ is a time-reversal even distribution, implying
$\textbf{h}^{\perp g}_{1,DIS}=\textbf{h}^{\perp g}_{1,DY}$. The linearly polarized gluon distribution
inside a large nucleus recently attracted a lot of attentions. $\textbf{h}^{\perp g}$ in the saturation regime was
first derived using the MV model in~\cite{Metz:2011wb}. Its rapidity evolution was also investigated~\cite{Dominguez:2011br}.
Many processes in which $\textbf{h}^{\perp g}$ can be probed have been proposed~\cite{Metz:2011wb,Dominguez:2011br,Schafer:2012yx,Liou:2012xy,Akcakaya:2012si}.

It is straightforward to extend our  analysis for  nuclear quark TMDs to  gluon TMDs. One thus obtains,
\begin{eqnarray}
\textbf{G}_{DIS}(x,k_\perp)&=& {\cal A} \int d^2 l_\perp  \, G_{DIS}(x,l_\perp) {\cal F}_{DIS}^g (|\vec k_\perp-\vec l_\perp|)
   \\
\textbf{h}_{1,DIS}^{\perp g}(x,k_\perp)&=& {\cal A} \int d^2 l_\perp \,\left [ 2 (\hat k_\perp \cdot \hat l_\perp)^2-1 \right ] \,
h_{1,DIS}^{\perp g} (x,l_\perp) {\cal F}_{DIS}^g (|\vec k_\perp-\vec l_\perp|) \ .
\end{eqnarray}
$G_{DIS}$ and $ h_{1,DIS}^{\perp g}$ are corresponding gluon
distributions in a nucleon. ${\cal F}_{DIS}^g (|\vec
k_\perp-\vec l_\perp|)$ is given by,
\begin{eqnarray}
 {\cal F}_{DIS}^g (|\vec k_\perp-\vec l_\perp|)=\int \frac{ d^2 r_{\perp}}{(2\pi)^2 }\, e^{ - i(\vec k_{\perp}-\vec l_\perp) \cdot \vec r_\perp} \,
4 \frac{1-e^{-\frac{r_\perp^2 Q_{s}^2}{4}}}{r_\perp^2 Q_{s}^2} \ ,
\end{eqnarray}
where $Q_{s}^2=\alpha_sN_c {\rm ln}\frac{1}{r_\perp^2 \Lambda_{QCD}^2}\int_{-\infty}^\infty d\zeta^- \, \lambda_A(\zeta^-)$
is the gluon saturation momentum. With these relations, it is easy to further verify that,
\begin{eqnarray}
\int d^2 k_\perp \, k_\perp^2 \textbf{G}_{DIS}(x,k_\perp)&=& {\cal A } \int d^2 l_\perp  \,  l_\perp^2 G_{DIS}(x,l_\perp)
+\frac{1}{2} Q_{s}^2 {\cal A} G_{DIS}(x)
  \\
\int d^2 k_\perp \,  k_\perp^2 \, \textbf{h}_{1,DIS}^{\perp g}(x,k_\perp)&=& {\cal A}  \int d^2 l_\perp \, l_\perp^2 \,
h_{1,DIS}^{\perp g}(x,l_\perp)  \ .
\end{eqnarray}
Correspondingly, the quark antiquark momentum imbalance in eA collisions is again of order $Q_{s}^2/2$.

\section{Nuclear  TMDs in photon-jet and heavy quark  pair production  in pA collisions }
When both initial and final state interactions are present in a hard scattering,
 a more complicated structure of gauge links
different from a simple past-pointing or future-pointing ones will appear.
This results in the process dependent nuclear $k_\perp$ broadening.

\subsection{Nuclear  TMDs in photon-jet production  }
We first study the nuclear enhancement of the transverse momentum imbalance for photon-jet production
in pA collisions,
\begin{eqnarray}
p(P')+A(P) \rightarrow \gamma(p_1)+{\rm Jet}(p_2)+X \, ,
\end{eqnarray}
where $P'$ and $P$ are the momenta of incoming proton and nucleus(per nucleon),
and $p_1$, $p_2$ are the momenta of the produced photon and jet respectively.
The transverse momentum imbalance $\vec q_\perp$ is  defined as: $\vec q_\perp=\vec p_{1\perp}+\vec p_{2\perp}$.

For the $q\bar q \rightarrow \gamma  g$ channel, the corresponding gauge link is built up by both initial state and
 final state interactions.  The  resulting nuclear quark TMDs are give by~\cite{Bomhof:2004aw},
\begin{eqnarray}
{\cal M}_A(x,\vec k_\perp) & = & \int \frac{d r^- d^2 r_{\perp}}{(2\pi)^3 }
\, e^{ixP^+r^- - i \vec k_{\perp} \cdot \vec r_\perp}
\nonumber\\ &&\times
\langle A | \bar \psi (y^-,y_\perp) \,
\left \{ \frac{9}{8N_c} U^{[+]} {\rm Tr}_c[U^{[ \!\!\!\!\!\!\!\! \qed ]\dag}]- \frac{1}{8} U^{[-]} \right \}
 \psi(r^-+ y^-, r_{\perp}+y_\perp )   |A \rangle
\nonumber\\ & = & \frac{1}{2} \textbf{f}_{1,q\bar q \rightarrow \gamma  g}(x,k_\perp)p \!\!\!/
+\frac{1}{2k_\perp}\textbf{h}^\perp_{1,q\bar q \rightarrow \gamma  g}(x,k_\perp) \sigma^{\mu\nu}k_\mu p_\nu
\,,
\label{photon}
\end{eqnarray}
where $U^{[ \!\!\!\!\!\!\!\! \qed ]}=U^{[+]}U^{[-]\dag}=U^{[-]\dag} U^{[+]}$ emerges as a Wilson loop.
The technique for evaluating multiple point correlation functions in the MV model has been systematically developed in Ref.~\cite{Blaizot:2004wv}.
The main strategy is to repeatedly use the Fierz identity $t^a_{ij}t^a_{kl}=\frac{1}{2}\delta_{il}\delta_{jk}-\frac{1}{2N_c}\delta_{ij}\delta_{kl}$
in order to resolve the color structure when gluon links connect different gauge links.
By closely following the  method presented in~\cite{Blaizot:2004wv},
we compute part of the gauge link $U^{[+]}U^{[+]\dag}$ in the MV model,
\begin{figure}[t]
\begin{center}
\includegraphics[width=17cm]{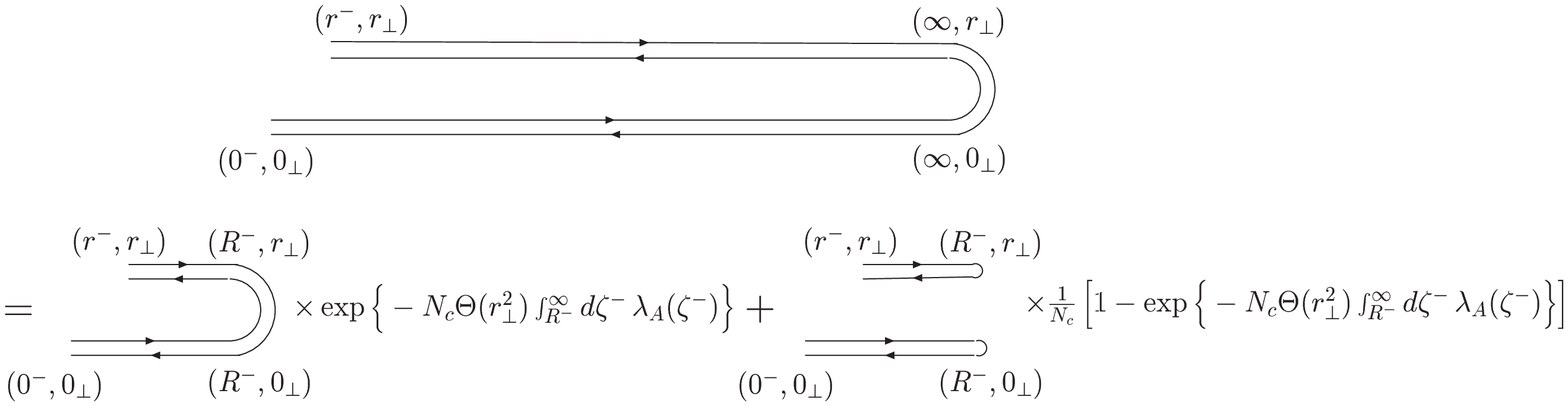}
\end{center}
\vskip -0.4cm \caption{The  gauge link $U^{[+]}U^{[+]\dag}$ is reduced to two short ones with different color structures
by evaluating part of the gauge link stretching from $R^-$ to $\infty$ in the MV model, where $R^-$ is the radius of a nucleon.} \label{fig1}
\end{figure}
\begin{eqnarray}
&& \!\!\!\!\!\!\!\!
\left <  \left [ \mathcal{P} \, e^{-ig \int_{y^-+R^-}^{\infty} d \zeta^- A^+(\zeta^-, \, y_{\perp})} \,
\mathcal{P} \, e^{-ig \int_\infty^{y^-+R^-}
d\zeta^-  A^+(r_\perp+y_\perp, \, \zeta^-)} \right ]_{ij} \right .\
\nonumber\\ && \left .\ \times
\left [ \mathcal{P} \, e^{-ig \int_{y^-+R^-}^{\infty} d \zeta^- A^+(\zeta^-, \, y_{\perp})} \,
\mathcal{P} \, e^{-ig \int_\infty^{y^-+R^-}
d\zeta^-  A^+(r_\perp+y_\perp, \, \zeta^-)} \right ]^\dag_{lm} \right >
\nonumber\\ &=&
\frac{1}{N_c}\left [1- e^ {  - N_c \Theta(r_\perp^2)\int_{R^-+y^-}^{\infty} d\zeta^- \, \lambda_A(\zeta^-)  } \right ]\delta_{im} \delta_{jl}+
e^ {  -N_c \Theta(r_\perp^2)\int_{R^-+y^-}^{\infty} d\zeta^- \, \lambda_A(\zeta^-)  } \delta_{ij} \delta_{lm} \, ,
\end{eqnarray}
where $i, j, l$ and $m$ are color indices.
It is worthwhile to mention that two different topologies show up as illustrated in Fig.2.
The gauge links $U^{[-]}$ and  $U^{[+]}$  have also been calculated in the previous section.
Inserting these results into Eq.\ref{photon}, one obtains,
\begin{eqnarray}
&& \!\!\!\!\!\!\!\!\!\!\!\!\!\!
{\cal M}_A(x,\vec k_\perp)  = \int \frac{d r^- d^2 r_{\perp}}{(2\pi)^3 }
\, e^{ixP^+r^- - i\vec{k}_{\perp} \cdot \vec{r}_\perp}   \int dy^- d^2y_\perp \,  \rho_N^A(y^-)
\nonumber \\ &\times& \!\!\!\!\!
\left \{
 \langle N | \bar \psi (0^-,0_\perp) \, \frac{9}{8} \bar U^{[+]} \frac{{\rm Tr}_c[ \bar U^{[ \!\!\!\!\!\!\!\! \qed ]\dag}]}{N_c} \, \psi(r^-, r_{\perp} )   |N \rangle
 e^{ - \Theta(r_\perp^2) \left [ C_F   \int^{y^--R^-}_{-\infty} d\zeta^- \, \lambda_A(\zeta^-)+ N_c \int_{y^-+R^-}^{\infty} d\zeta^- \, \lambda_A(\zeta^-) \right ] }
  \right .\
\nonumber \\ &+& \!\!\!
 \langle N | \bar \psi (0^-,0_\perp) \, \frac{1}{8}\bar U^{[-]} \, \psi(r^-, r_{\perp} )   |N \rangle
e^{ - C_F \Theta(r_\perp^2) \int^{y^--R^-}_{-\infty} d\zeta^- \, \lambda_A(\zeta^-) }
\left [ 1 -e^{ - N_c \Theta(r_\perp^2) \int^{\infty}_{y^-+R^-} d\zeta^- \, \lambda_A(\zeta^-) } \right ]
\nonumber \\ &-&   \left .\ \!\!\!\!\!\!
 \langle N | \bar \psi (0^-,0_\perp) \, \frac{1}{8}\bar U^{[-]} \, \psi(r^-, r_{\perp} )   |N \rangle
e^{ - C_F \Theta(r_\perp^2) \int^{y^--R^-}_{-\infty} d\zeta^- \, \lambda_A(\zeta^-) } \right \} \, .
\end{eqnarray}
The short Wilson loop appearing in the above equation consists of a short future-pointing and a short past-pointing gauge link:
$\bar U^{[ \!\!\!\!\!\!\!\! \qed ]}=\bar U^{[+]}\bar U^{[-]\dag}$.
By approximating the  large nucleus as a homogenous system of color sources, we are able to carry out the
integration over $y^-$ and $y_\perp$. One then ends up with,
\begin{eqnarray}
&& \!\!\!\!\!\!\!\!\!\!\!\!\!\!
{\cal M}_A(x,\vec k_\perp)  =  \int \frac{d r^- d^2 r_{\perp}}{(2\pi)^3 }
\, e^{ixP^+r^- - i\vec{k}_{\perp} \cdot \vec{r}_\perp}
\nonumber \\ && \times
 {\cal A} \langle N | \bar \psi (0^-,0_\perp) \, \left \{ \frac{9}{8} \bar U^{[+]} \frac{{\rm Tr}_c[ \bar U^{[ \!\!\!\!\!\!\!\! \qed ]\dag}]}{N_c}
- \frac{1}{8}\bar U^{[-]} \right \} \, \psi(r^-, r_{\perp} )   |N \rangle
 \left [   \frac{e^{\frac{-Q_s^2r_\perp^2}{4}}-e^{-\frac{Q_{s,q}^2r_\perp^2}{4}}}{(Q_{s,q}^2-Q_{s}^2)r_\perp^2/4} \right ] \, ,
\end{eqnarray}
where $Q_{s}^2=\alpha_s N_c {\rm ln}\frac{1}{r_\perp^2 \Lambda_{QCD}^2}\int_{-\infty}^\infty d\zeta^- \, \lambda_A(\zeta^-)$
is the gluon saturation momentum. To arrive the above equation, we have made one further approximation,
 $\int^{y^--R^-}_{-\infty} d\zeta^- \, \lambda_A(\zeta^-) +\int_{y^-+R^-}^{\infty} d\zeta^- \, \lambda_A(\zeta^-)
\approx \int_{-\infty}^{\infty} d\zeta^- \, \lambda_A(\zeta^-)$, which is valid for a large nucleus.
One can readily transform this expression to  momentum space,
\begin{eqnarray}
\textbf{f}_{1,q\bar q \rightarrow \gamma  g}(x,k_\perp)&=& {\cal A}
\int d^2 l_\perp  \, f_{1,q\bar q \rightarrow \gamma  g}(x,l_\perp)
{\cal F}_{q\bar q \rightarrow \gamma  g}(|\vec k_\perp-\vec l_\perp|)
   \\
\textbf{h}_{1,q\bar q \rightarrow \gamma  g}^\perp(x,k_\perp)&=& {\cal A}
\int d^2 l_\perp \, (\hat k_\perp \cdot \hat l_\perp) \,
h_{1,q\bar q \rightarrow \gamma  g}^\perp(x,l_\perp) {\cal F}_{q\bar q \rightarrow \gamma  g}(|\vec k_\perp-\vec l_\perp|) \ ,
\end{eqnarray}
with ${\cal F}_{q\bar q \rightarrow \gamma  g}(k_\perp)$ being given by,
\begin{eqnarray}
 {\cal F}_{q\bar q \rightarrow \gamma  g}(|\vec k_\perp-\vec l_\perp|)=
 \int \frac{ d^2 r_{\perp}}{(2\pi)^2 }\, e^{ - i(\vec k_\perp-\vec l_\perp) \cdot \vec r_\perp} \,
\frac{e^{\frac{-Q_s^2r_\perp^2}{4}}-e^{\frac{-Q_{s,q}^2r_\perp^2}{4}}}{(Q_{s,q}^2-Q_{s}^2)r_\perp^2/4} \ .
\end{eqnarray}
Correspondingly, the $k_\perp$ moment of nuclear TMDs read,
\begin{eqnarray}
\int d^2 k_\perp \, k_\perp^2 \textbf{f}_{1,q\bar q \rightarrow \gamma  g}(x,k_\perp)&=&
 {\cal A} \int d^2 l_\perp  \,  l_\perp^2 f_{1,q\bar q \rightarrow \gamma  g}(x,l_\perp)
 +\left [ \frac{1}{2} Q_{s,q}^2+ \frac{1}{2} Q_{s}^2\right  ]{\cal A} f_{1}(x)
  \\
\int d^2 k_\perp \,  k_\perp \, \textbf{h}_{1,q\bar q \rightarrow \gamma  g}^\perp(x,k_\perp)&=&
 {\cal A}  \int d^2 l_\perp \,  l_\perp \, h_{1,q\bar q \rightarrow \gamma  g}^\perp(x,l_\perp)=
-2\pi M_N {\cal A} \frac{N_c^2+1}{N_c^2-1} T_F^{(\sigma)}(x,x)  \ ,
\end{eqnarray}
where the nontrivial color factor $(N_c^2+1)/(N_c^2-1)$ originates from the T-odd nature of the quark Boer-Mulders distribution.
In the semihard region where the imbalance of the photon-jet produced in pA collisions is of the order  $Q_s\gg\Lambda_{QCD}$,
 after neglecting the terms suppressed by the power of $\Lambda_{QCD}^2/Q_s^2$, we have,
\begin{eqnarray}
\textbf{f}_{1,q\bar q \rightarrow \gamma  g}(x,k_\perp)&\simeq& {\cal A}
 f_{1}(x) {\cal F}_{q\bar q \rightarrow \gamma  g}(k_\perp)
   \\
\textbf{h}_{1,q\bar q \rightarrow \gamma  g}^\perp(x,k_\perp)&\simeq&
{\cal A}  2\pi M_N \frac{N_c^2+1}{N_c^2-1} T_F^{(\sigma)}(x,x)
\frac{1}{2} \frac{\partial {\cal F}_{q\bar q \rightarrow \gamma  g}(k_\perp)}{\partial k_\perp}
 \ .
\end{eqnarray}

The nuclear quark Boer-Mulders function can manifest itself through $\cos 2\phi$ asymmetries as can be seen by convoluting
with the function $T_F^{(\sigma)}$ from the proton side.
Moreover, if the incoming proton is transversely polarized,
the quark Boer-Mulders function can  couple with the transversity distribution of the proton and
give rise to the single transverse spin asymmetry(SSA).
Such observable in pA collisions was also studied  from a different points of view in the papers~\cite{Boer:2002ij,Kang:2011ni,Kovchegov:2012ga,Kang:2012vm}.

Photon-jet pair can also be produced through other three channels:
 $ \bar  q q\rightarrow \gamma  g$, $qg \rightarrow \gamma  q$ and $g q \rightarrow \gamma  q$.
 The associated nuclear TMDs contain different gauge link structures in the different channels.
 By evaluating the gauge links in the MV model, it is straightforward to establish
  the  relations between nucleon TMDs and nuclear TMDs in these processes.
 However, for simplicity, we only list the $k_\perp$ momenta of the corresponding nuclear TMDs,
\begin{eqnarray}
\int d^2 k_\perp \, k_\perp^2 \bar \textbf{f}_{1,\bar q q \rightarrow \gamma  g}(x,k_\perp)&=&
 {\cal A} \int d^2 l_\perp  \,  l_\perp^2 \bar f_{1,\bar q q \rightarrow \gamma  g}(x,l_\perp)
 +\left [ \frac{1}{2} Q_{s,q}^2+ \frac{1}{2} Q_{s}^2\right  ] {\cal A}  \bar f_{1}(x)
  \\
\int d^2 k_\perp \, k_\perp^2 \textbf{f}_{1,qg \rightarrow \gamma  q}(x,k_\perp)&=&
  {\cal A}  \int d^2 l_\perp  \,  l_\perp^2 f_{1,qg \rightarrow \gamma  q}(x,l_\perp)
 +\left [ \frac{1}{2} Q_{s,q}^2+ \frac{1}{2} Q_{s}^2\right  ] {\cal A} f_{1}(x)
  \\
\int d^2 k_\perp \, k_\perp^2 \textbf{G}_{g q \rightarrow \gamma  q}(x,k_\perp)&=&
 {\cal A}  \int d^2 l_\perp  \,  l_\perp^2 G_{g q \rightarrow \gamma  q}(x,l_\perp)
 +  Q_{s,q}^2 {\cal A}  G(x) \, ,
\end{eqnarray}
where  $\bar \textbf{f}_{1}$ denotes the anti-quark distribution.

The average  squared transverse momentum imbalance is given by,
\begin{eqnarray}
<q_\perp^2> = \left ( \int d^2q_\perp q_\perp^2 \frac{d\sigma}{d {\cal P.S.} d^2 q_\perp}  \right ) /
\frac{d\sigma}{d {\cal P.S.} }
\end{eqnarray}
where $d {\cal P.S.}=dy_1 dy_2 d^2 p_\perp$ stands for the phase space with $y_1$, $y_2$ being the produced photon and
jet rapidities respectively. The nuclear broadening of the photon-jet imbalance is defined as
\begin{eqnarray}
\Delta \!\! <q_\perp^2>= <q_\perp^2>_{pA}-<q_\perp^2>_{pp} \, ,
\end{eqnarray}
Putting all these together, within the TMD factorization framework the  nuclear enhancement of the photon-jet
squared transverse momentum imbalance is given by,
\begin{eqnarray}
\Delta \!\! <q_\perp^2>=\Delta \!\! <k_\perp^2>=\frac{1}{2} Q_{s,q}^2+ \frac{1}{2} Q_{s}^2-
\frac{ \left [ \frac{1}{2} Q_s^2- \frac{1}{2} Q_{s,q}^2 \right ]
\Sigma_a  H_{gq \rightarrow\gamma q}^a  }
{\Sigma_a \left [H_{qg\rightarrow\gamma q}^a+H_{gq \rightarrow\gamma q}^a+H_{q \bar q \rightarrow\gamma g}^a+ H_{\bar q  q \rightarrow\gamma g}^a \right ]} \, ,
\end{eqnarray}
where $a$ runs all quark flavors.
The partonic hard scattering differential cross sections read~\cite{Owens:1986mp},
\begin{eqnarray}
H_{qg\rightarrow\gamma q}^a&=&e_q^2 \frac{1}{N_c} \left ( -\frac{\hat s}{\hat t}-\frac{\hat t}{\hat s} \right )
\textbf{f}^a_1(x) G(x')
\\
H_{gq \rightarrow\gamma q}^a&=&e_q^2 \frac{1}{N_c} \left ( -\frac{\hat s}{\hat u}-\frac{\hat u}{\hat s}  \right )
\textbf{G}(x) f^a_1(x')
\\
H_{q \bar q \rightarrow\gamma g}^a&=& e_q^2 \frac{N_c^2-1}{N_c^2} \left ( \frac{\hat t}{\hat u}+\frac{\hat u}{\hat t}  \right )
\textbf{f}^a_1(x) \bar f^a_1(x')
\\
H_{\bar q  q \rightarrow\gamma g}^a&=& e_q^2 \frac{N_c^2-1}{N_c^2} \left ( \frac{\hat t}{\hat u}+\frac{\hat u}{\hat t}  \right )
 \bar \textbf{f}^a_1(x)  f^a_1(x')
\end{eqnarray}
Here, $\hat s$, $\hat u$ and $\hat t$ are the usual partonic Mandelstam variables.
The collinear momentum fraction is fixed by the kinematical constraint,
\begin{eqnarray}
x'=\frac{p_\perp}{\sqrt{s}}(e^{y_1}+e^{y_2}) \   ,  \ x'=\frac{p_\perp}{\sqrt{s}}(e^{-y_1}+e^{-y_2})
\end{eqnarray}
where $s=(P+P')^2$ is the center of mass energy squared. By noticing $Q_s^2=\frac{C_A}{C_F} Q_{s,q}^2$,
the nuclear broadening of the photon-jet imbalance varies from $1\frac{5}{8} \, Q_{s,q}^2$ to $ Q_{s,q}^2$ depending on the specific kinematical variables.

\subsection{$k_\perp$ broadening in heavy quark pair production }
Quark antiquark pair production in high energy pA collisions is dominated by the
gluon  initiated parton subprocess,
\begin{eqnarray}
g_p(x_1P')+g_A(x_2P) \rightarrow Q(p_1)+\bar Q(p_2) \, .
\end{eqnarray}
The nuclear enhancement of the quark pair transverse momentum imbalance $\vec q_\perp=\vec p_{1\perp}+\vec p_{2\perp}$
is directly sensitive to nuclear gluon TMD distributions. The color flow in this subprocess is more complicated than
that for  photon-jet production in pA collisions.
The nuclear gluon TMDs associated with different Feynman diagrams contain different gauge link structures. Here we show two examples,
\begin{eqnarray}
{\cal M}_A^a &\propto &
\langle A | {\rm Tr_c} \left [ F_{+\perp}  \,
\left \{ \frac{9}{8} \frac{ {\rm Tr}_c[U^{[ \!\!\!\!\!\!\!\! \qed ]\dag}]}{N_c}U^{[-]\dag}
- \frac{1}{8} U^{[+]^\dag} \right \}
 F_{+\perp}   U^{[+]}\right ] |A \rangle
\\
{\cal M}_A^e &\propto &
\langle A | {\rm Tr_c}
\left [ F_{+\perp} U^{[-]\dag}F_{+\perp} U^{[+]} \right ]
\frac{ {\rm Tr}_c[U^{[ \!\!\!\!\!\!\!\! \qed ]\dag}]}{N_c} \,
- \frac{1}{N_c} {\rm Tr_c} \left [
{\rm F_{+\perp}}U^{[ \!\!\!\!\!\!\!\! \qed ]\dag} \right ]
 {\rm Tr_c} \left [ {\rm F_{+\perp}} U^{[ \!\!\!\!\!\!\!\! \qed ]} \right ]
 |A \rangle \, ,
\end{eqnarray}
where gauge links appearing in ${\cal M}_A^a$ and ${\cal M}_A^e$ originate from initial/final state interactions
in the Feynman diagrams Fig.3(a) and Fig.3(e), respectively.
All other gauge links appearing in this hard scattering process  are given in~\cite{Bomhof:2004aw}.
One can compute the transverse momentum spectrum of the gluon distribution associated with each
Feynman diagram following a similar method as introduced in the previous subsections.  With the derived nuclear gluon
TMDs, we obtain,
\begin{eqnarray}
\int d^2 k_\perp \, k_\perp^2 \textbf{G}_{g g \rightarrow Q \bar Q}^a(x,k_\perp)&=&
 {\cal A}  \int d^2 l_\perp  \,  l_\perp^2 G_{g g \rightarrow Q \bar Q}^a(x,l_\perp)
 +  \left [ 1-\frac{1}{2}\frac{1}{N_c^2-1} \right ] Q_{s}^2 {\cal A}  G(x)
\\
\int d^2 k_\perp \, k_\perp^2 \textbf{G}_{g g \rightarrow Q \bar Q}^b(x,k_\perp)&=&
 {\cal A}  \int d^2 l_\perp  \,  l_\perp^2 G_{g g \rightarrow Q \bar Q}^b(x,l_\perp)
 +  \left [ 1-\frac{1}{2}\frac{1}{N_c^2-1} \right ] Q_{s}^2 {\cal A}  G(x)
\\
\int d^2 k_\perp \, k_\perp^2 \textbf{G}_{g g \rightarrow Q \bar Q}^c(x,k_\perp)&=&
 {\cal A}  \int d^2 l_\perp  \,  l_\perp^2 G_{g g \rightarrow Q \bar Q}^c(x,l_\perp)
 +  \frac{3}{2} Q_{s}^2 {\cal A}  G(x)
\\
\int d^2 k_\perp \, k_\perp^2 \textbf{G}_{g g \rightarrow Q \bar Q}^d(x,k_\perp)&=&
 {\cal A}  \int d^2 l_\perp  \,  l_\perp^2 G_{g g \rightarrow Q \bar Q}^d(x,l_\perp)
 +  Q_{s}^2 {\cal A}  G(x)
\\
\int d^2 k_\perp \, k_\perp^2 \textbf{G}_{g g \rightarrow Q \bar Q}^e(x,k_\perp)&=&
 {\cal A}  \int d^2 l_\perp  \,  l_\perp^2 G_{g g \rightarrow Q \bar Q}^e(x,l_\perp)
 +  Q_{s}^2 {\cal A}  G(x)
\\
\int d^2 k_\perp \, k_\perp^2 \textbf{G}_{g g \rightarrow Q \bar Q}^f(x,k_\perp)&=&
 {\cal A}  \int d^2 l_\perp  \,  l_\perp^2 G_{g g \rightarrow Q \bar Q}^f(x,l_\perp)
 +  Q_{s}^2 {\cal A}  G(x) \, .
\end{eqnarray}
It is easy to verify that the hard coefficient computed from the Feynman diagram Fig.3(c) is suppressed by the factor of $1/N_c^2$
as compared to the contributions from other diagrams in the large $N_c$ limit.
Thus, we conclude that the nuclear $k_\perp$ broadening for quark anitquark pair production
in pA collisions is $Q_s^2$ in the large $N_c$ limit.

\begin{figure}[t]
\begin{center}
\includegraphics[width=11cm]{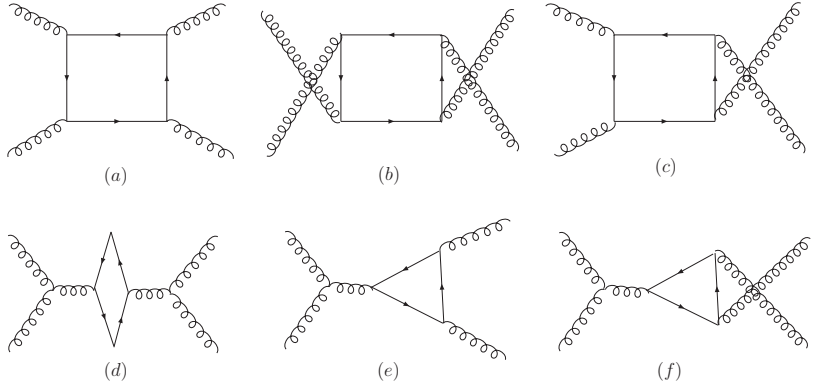}
\end{center}
\vskip -0.4cm \caption{Feynman diagrams contributing to quark antiquark pair production.} \label{fig3}
\end{figure}

\section{Phenomenology applications  }
In this section, we discuss  phenomenology applications of our results and compare our formalism with the higher twist
collinear approach.  We start from numerically evaluating  nuclear $k_\perp$ broadening for jet production in eA collisions.
The analytical result for this observable takes a simple form,
\begin{eqnarray}
\Delta<k_\perp^2>_{\gamma^* q\rightarrow q }&=&Q_{s,q}^2/2 \ .
\end{eqnarray}
However, in deriving the above result, we only took into account the contributions to the gauge link from color sources outside of the nucleon to which
the struck parton belongs. To remedy this problem, the above equation should be slightly modified as follows,
\begin{eqnarray}
\Delta<k_\perp^2>_{\gamma^* q\rightarrow q }&=&\frac{{\cal A}^{1/3}-1}{{\cal A}^{1/3}}Q_{s,q}^2/2 \ ,
\end{eqnarray}
where we adopt the parametrization for the saturation momentum used in the GBW model~\cite{GolecBiernat:1998js}:
 $Q_s^2={\cal A}^{1/3} Q_0^2(x_0/x_g)^\lambda $ with $Q_0^2=1 GeV^2$, $x_0=3\cdot 10^{-4}$ and $\lambda\approx 0.3$.
 For a lead or gold targets, ${\cal A}^{1/3}-1$ is approximately equal to  $5$.

The natural next step is to fix $x_g$. At first glance, gluons building up gauge links carry exactly zero longitudinal
momentum due to the contour integration around the gluon pole $1/(x_g-i\epsilon)$(for  final state interactions).
This pole arises when performing the calculation in the collinear approximation.
However, if one keeps the gluon transverse momentum $k_{g\perp}$, the gluon pole in jet production in the SIDIS process will be modified to,
\begin{eqnarray}
\frac{1}{x_g-k_{g\perp}^2/(2p \cdot q) -i\epsilon} \, ,
\end{eqnarray}
where $q$ is the virtual photon momentum.
Within the  leading logarithm accuracy,
it is convenient to replace $k_{g\perp}^2$ in  above formula by
$\Delta \!\! <k_{\perp}^2>$.
Once $x_g$ is fixed as $\Delta <k_{\perp}^2>/(2p \cdot q)$, one obtains (for a lead or gold target),
\begin{eqnarray}
\Delta <k_{\perp}^2>_{\gamma^* q\rightarrow q }=
\left [\frac{1}{2} \frac{C_F}{C_A} \left ({\cal A}^{1/3}-1\right )Q_0^2 \right ]^{1/(1+\lambda)}
\left ( s \frac{x_0}{1+x_B} \right )^{\lambda/(1+\lambda)}
\approx 1.08 \left ( s \frac{x_0}{1+x_B} \right )^{0.23}\left ( Q_0^2 \right )^{0.77} \, ,
\end{eqnarray}
where $s=(p+q)^2$ and  $x_B\equiv Q^2/2p\cdot q=Q^2/(s+Q^2)$  with $Q^2=-q^2$ being the virtual photon virtuality.
The above equation holds as long as $x_B$ is of the order of 1
 and $x_g \leq 0.01$ where the MV model can apply.
 This implies that our formalism is invalid at relatively low energy~\cite{Airapetian:2009jy}.
Given $x_B=0.2$ and $\sqrt s=35GeV$ accessible at a future EIC, the nuclear $k_\perp$
broadening for jet production in the SIDIS process gives
$\Delta \!\! <k_{\perp}^2>_{\gamma^* q\rightarrow q } \, \approx 0.82GeV^2$.
For the Drell-Yan process one can fix the saturation scale in the same way.
Unfortunately, we are not able to  unambiguously determine the saturation scale for other processes.

Now we compare our result with that obtained from the  higher twist collinear approach.
As mentioned in the introduction, the process dependent $k_\perp$ broadening effect was also investigated within
the higher twist collinear factorization framework~\cite{Kang:2011bp,Xing:2012ii}
 which can be applied in the  intermediate or large $x$ region.
In this formalism, the effect of initial/final state multiple scattering generating $k_\perp$ broadening is
encoded in the collinear twist-4 quark-gluon correlation functions $T^{(I)}_{q,g/A}(x)$ and $T^{(F)}_{q,g/A}(x)$.
These functions are parameterized as follows,
\begin{eqnarray}
\frac{4\pi^2 \alpha_s}{N_c} T^{(I)}_{q,g/A}(x)=\frac{4\pi^2 \alpha_s}{N_c} T^{(F)}_{q,g/A}(x)
=\xi^2 ({\cal A}^{1/3}-1)f_{q,g/A} (x) \, ,
\end{eqnarray}
where $\xi^2$ represents a characteristic  scale of parton multiple scattering,
  and $f_{q,g/A} (x)$ is the standard leading-twist parton distribution function for quarks and gluons, respectively.
If we identify  the saturation scale as,
\begin{eqnarray}
\frac{1}{2} \frac{Q^{2}_{s,q}}{C_F} = \frac{1}{2} \frac{Q^{2}_{s}}{C_A}
=\xi^2 {\cal A}^{1/3} \, ,
\label{param}
\end{eqnarray}
our analytical results for $k_\perp$ broadening take the same form as those presented in~\cite{Kang:2011bp,Xing:2012ii}.
In~\cite{Kang:2011bp,Xing:2012ii}, $\xi^2$ is chosen to be $0.12 GeV^2$. Provided that one evaluates the saturation scale using the same value
$\xi^2=0.12 GeV^2$, we have numerical result for jet $k_\perp$ broadening in SIDIS
 $\Delta \!\! <k_{\perp}^2>_{\gamma^* q\rightarrow q } \, = 0.8 GeV^2$, which is very close to that calculated with
 the GBW parametrization. We  would also get the identical  numerical results for all other processes using Eq.\ref{param}.

Apart from  the process dependent unpolarized nuclear TMDs,
 the process dependent nuclear quark Boder-Mulders function is also studied in this paper.
Below we discuss the corresponding phenomenological implications for the polarized cases.
In both SIDIS  and DY  in pA collisions,
the nuclear quark Boer-Mulders function can give rise to  $\cos 2 \phi$ azimuthal asymmetries by coupling with
the Collins fragmentation function and the anti-quark Boer-Mulders function from the proton side\cite{Boer:1999mm},
respectively. According to our calculation, in the semi-hard region,
the transverse momentum dependence of the asymmetry is unambiguously  determined by the ratio,
$$<\cos 2 \phi>\!\!(k_\perp) \  \propto \frac{\partial {\cal F}_{DIS}(k_\perp)}{\partial k_\perp}  /  {\cal F}_{DIS}(k_\perp) \, .$$
Such observables can in principle  be measured in unpolarized $eA$ collisions at  EIC and unpolarized pA collisions at RHIC.
 We leave  detailed phenomenological studies for a future work.

\section{Summary}
We have established relations between nuclear TMDs and the corresponding nucleon ones
by computing contributions from the process dependent gauge links in the MV model.
 In particular, in the semi-hard region where quark transverse momenta are of the order of the saturation scale,
unpolarized nuclear TMDs are determined by the process dependent small $x$ gluon distributions, while nuclear quark
Boer-Mulders distributions are expressed as  products of $T_F^{(\sigma)}(x,x)$ and the differential of the same gluon distributions with respect to
gluon transverse momentum. We stress again that the formalism developed in this paper applies only for nuclear TMDs at
intermediate or large $x$.

Two phenomenological applications of our work are nuclear $k_\perp$ broadening and the $k_\perp$ dependence of the asymmetries
generated by the quark Boer-Mulders function in eA and pA collisions.  To be more specific, we calculated  nuclear $k_\perp$
broadening for jet and di-jet production in eA collisions, and the nuclear enhancement of the transverse momentum imbalance
for Drell-Yan lepton pair production, photon-jet production, and quark antiquark pair production  in  pA collisions.
 To investigate how the quark Boer-Mulders function are affected by the surrounding cold nuclear matter,
we also proposed to measure $\cos 2\phi$ azimuthal asymmetries in SIDIS off a large nucleus and Drell-Yan pairs in pA collisions.
It will be interesting to test our predications at RHIC and the planned EIC.
%
%

\

\noindent
{\bf Acknowledgments:}
This work has been supported by BMBF (OR 06RY9191 and 05P12WRFTE).
%
%

\end{document}